\documentclass[11pt]{article}
\usepackage{moriond}
\usepackage{epsfig}
\usepackage[latin1]{inputenc}
\usepackage[english]{babel}
\usepackage{array}
\usepackage{amssymb}
\usepackage{amsmath}
\usepackage{feynmf}
\usepackage{dcolumn}
\usepackage{color}
\definecolor{brown}{rgb}{0.5,0.4,0.2}
\definecolor{lightblue}{rgb}{0.5,0.5,1.0}
\definecolor{darkgreen}{cmyk}{0.5, 0, 1, 0.5}
\definecolor{darkgrey}{rgb}{0.2,0.2,0.2}
\definecolor{lightgrey}{rgb}{0.6,0.6,0.6}

\newcolumntype{d}[1]{D{.}{.}{#1}}

\newcommand{\ee}{\mbox{\ensuremath{{\mathrm{e}}^+ {\mathrm{e}}^-}}}

\def\be{\begin{equation}}
\def\ee{\end{equation}}
\def\bea{\begin{eqnarray}}
\def\eea{\end{eqnarray}}

\begin{document}
\begin{fmffile}{fmftempl}

%
%

\vspace*{4cm}
\title{Radiative \boldmath$b\to d$\unboldmath{} Penguins}

\author{ P. Bechtle~\footnote{for the BaBar collaboration} }

\address{Stanford Linear Accelerator Center, 2575 Sand Hill Road, Menlo Park, CA 94025, USA~\footnote{now: DESY, Notkestra\ss{}e 85, 22607 Hamburg, Germany, \texttt{philip.bechtle@desy.de}}}

\maketitle

\begin{abstract} 
  This article gives an overview of the recent searches and
  measurements of $b\to d$ penguin transitions with the BaBar
  experiment. The branching fraction of these decays in the Standard
  Model (SM) is expected to be a factor of 10 or more lower than the
  corresponding $b\to s$ penguin transitions, but a deviation from the
  SM prediction would be an equally striking sign of new physics. The
  exclusive decay $B\to\pi\ell\ell$ is searched by BaBar with no
  excess over the background found. The BaBar measurement of
  $B\to(\rho,\omega)\gamma$ provides the first evidence of
  $B^+\to\rho^+\gamma$,~\footnote{charge conjugation is implied
    everywhere in this paper, if not stated otherwise} is in good
  agreement with the previous Belle results and provides a measurement
  of $|V_{td}/V_{ts}|$ independent of the one from $B_s$ mixing. No
  deviation from the SM is found.
\end{abstract}

\section{Introduction}\label{sec:intro}

Since the first $\approx 10$ radiative penguin decays of B mesons have
been reported by CLEO in 1993~\cite{Ammar:1993sh}, the measurements of
radiative penguins at the B factory experiments BaBar and Belle have
expanded into a rich field of physics. As many as 15 individual
exclusive $b\to s$ modes have been observed and some of the
corresponding branching fractions have been measured with a precision
of about $10\,\%$. This makes it possible to test SM predictions of
these rare decays, e.g.{} QCD calculations of form factors.
Additionally, the inclusive measurement of $b\to s\gamma$ allows
strong constraints on new physics models~\cite{Misiak:2006zs} and
precision measurements of heavy quark effective theory predictions.

With the success of the $b\to s$ penguin measurements in hand, the
next step is to explore $b\to d$ penguins, which due to Cabbibo
suppression are not only about one order of magnitude more rare,
but also face a much more severe background due to the abundance of
pions in the background. This paper gives an overview of the recent
results from BaBar for the exclusive modes $B\to \pi\ell\ell$ and
$B\to(\rho,\omega)\gamma$.  Section~\ref{sec:newphysics} will briefly
discuss the interest in these measurements from the point of view of
the search for new physics, Sections~\ref{sec:btodll} and
\ref{sec:btodg} will summarize the measurements, an discuss the
measurement of $|V_{td}/V_{ts}|$ from ${\cal
  B}(B\to(\rho,\omega)\gamma)$.
 

\section{New Physics in Radiative Penguins}\label{sec:newphysics}

\begin{figure}[t]
  \begin{center}
    \begin{fmfgraph*}(20,15)
      \fmfstraight
      \fmfleft{throughin,bin,dummy1}\fmflabel{$\mathrm{b}$}{bin} \fmflabel{$\bar{\mathrm{u}}$}{throughin}
      \fmfright{throughout,sout,photonout}\fmflabel{$\mathrm{s,d}$}{sout}\fmflabel{$\gamma$}{photonout}
      \fmflabel{$\bar{\mathrm{u}}$}{throughout}
      \fmf{fermion,tension=2}{bin,bWl}
      \fmf{fermion,tension=2}{sWr,sout}
      \fmf{phantom}{bWl,sWr}
      \fmf{fermion}{throughin,throughout}
      \fmffreeze
      \fmf{photon,left=0.7,label=$\mathrm{W}^-$}{bWl,photonW}
      \fmf{photon,left=0.4}{photonW,sWr}
      \fmf{fermion,right,label=$\mathrm{t}$,label.side=left}{bWl,sWr}
      \fmf{photon,tension=1.5}{photonW,photonout}
    \end{fmfgraph*}\qquad\qquad
    \begin{fmfgraph*}(20,15)
      \fmfstraight
      \fmfleft{throughin,bin,dummy1}\fmflabel{$\mathrm{b}$}{bin}\fmflabel{$\bar{\mathrm{u}}$}{throughin}
      \fmfright{throughout,sout,photonout}\fmflabel{$\mathrm{s,d}$}{sout}\fmflabel{$\gamma$}{photonout}
      \fmflabel{$\bar{\mathrm{u}}$}{throughout}
      \fmf{fermion,tension=2}{bin,bWl}
      \fmf{fermion,tension=2}{sWr,sout}
      \fmf{fermion}{throughin,throughout}
      \fmf{phantom}{bWl,sWr}
      \fmffreeze
      \fmf{photon,left=0.7,label=$\chi_1^-$}{bWl,photonW}
      \fmf{photon,left=0.4}{photonW,sWr}
      \fmf{plain,right,label=$\tilde{\mathrm{t}}$,label.side=left}{bWl,sWr}
      \fmf{photon,tension=1.25}{photonW,photonout}
    \end{fmfgraph*}\qquad\qquad
    \begin{fmfgraph*}(20,15)
      \fmfstraight
      \fmfleft{throughin,bin,dummy1}\fmflabel{$\mathrm{b}$}{bin}\fmflabel{$\bar{\mathrm{u}}$}{throughin}
      \fmfright{throughout,sout,photonout}\fmflabel{$\mathrm{s,d}$}{sout}\fmflabel{$\gamma$}{photonout}
      \fmflabel{$\bar{\mathrm{u}}$}{throughout}
      \fmf{fermion,tension=2}{bin,bWl}
      \fmf{fermion,tension=2}{sWr,sout}
      \fmf{fermion}{throughin,throughout}
      \fmf{phantom}{bWl,sWr}
      \fmffreeze
      \fmf{dashes,left=0.7,label=$\mathrm{H}^-$}{bWl,photonW}
      \fmf{dashes,left=0.33}{photonW,sWr}
      \fmf{fermion,right,label=$\mathrm{t}$,label.side=left}{bWl,sWr}
      \fmf{photon,tension=1.25}{photonW,photonout}
    \end{fmfgraph*}
  \end{center}
  \caption{\sl{New Physics in Radiative Penguins. The left graph shows
      the SM contribution, the middle graph a possible SUSY
      contribution at the same loop level and with similar order of
      magnitudes of the couplings and masses. The right graph shows a
      SUSY or 2HDM contribution.}}
  \label{fig:newPhysicsGraphs}
\end{figure}
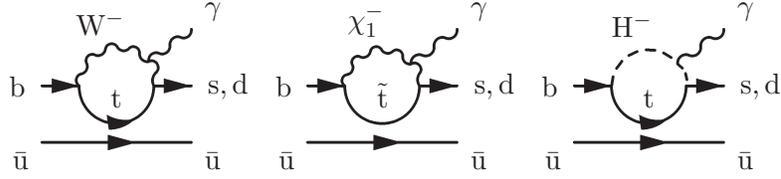

The beauty of the search for new physics in penguin decays lies in the
fact that the dominating diagram in the SM is a loop or box diagram
with heavy particles in the loop. This is exemplified in
Figure~\ref{fig:newPhysicsGraphs}. Possible new physics, Supersymmetry
(SUSY) or the two-Higgs Doublet Model (2HDM) in this case, contribute
with particles in the same mass range of 100 to several hundred GeV
and often also with similar couplings. This allows strong constraints
on new physics models. However, while the measurement of exclusive
decays is often experimentally clean, the interpretation of the
experimental results for very rare exclusive decays often suffers from
theoretical uncertainties in the prediction of form factors, radiative
corrections, and other suppressed graphs. Still, strong constraints
and precise SM measurements can be gained from ratios of branching
fractions.

\section[$b\to d\,\ell^+\ell^-$ Transitions]{\boldmath$b\to d\,\ell^+\ell^-$\unboldmath{} Transitions}\label{sec:btodll}

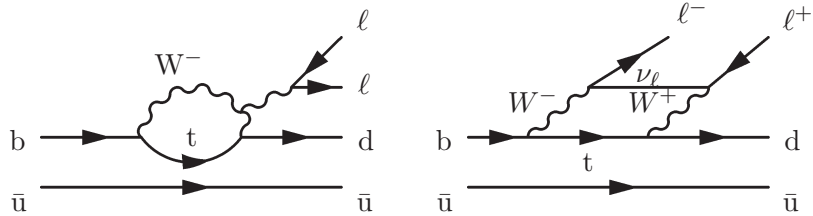
\begin{figure}[t]
  \vspace*{3mm}
  \begin{center}
    \begin{fmfgraph*}(40,20)
      \fmfstraight
      \fmfleft{throughin,bin,dummy1,dummy2}\fmflabel{$\mathrm{b}$}{bin} \fmflabel{$\bar{\mathrm{u}}$}{throughin}
      \fmfright{throughout,sout,leptonOneOut,leptonTwoOut}
      \fmflabel{$\mathrm{d}$}{sout}\fmflabel{$\ell$}{leptonOneOut}\fmflabel{$\ell$}{leptonTwoOut}
      \fmflabel{$\bar{\mathrm{u}}$}{throughout}
      \fmf{fermion,tension=1}{bin,bWl}
      \fmf{fermion,tension=1}{sWr,sout}
      \fmf{phantom}{bWl,sWr}
      \fmf{fermion}{throughin,throughout}
      \fmffreeze
      \fmf{photon,left=0.7,label=$\mathrm{W}^-$}{bWl,photonW}
      \fmf{photon,left=0.4}{photonW,sWr}
      \fmf{fermion,right=0.5,label=$\mathrm{t}$,label.side=left}{bWl,sWr}
      \fmf{photon,tension=2}{photonW,photonout}
      \fmf{fermion,tension=1}{leptonTwoOut,photonout}
      \fmf{fermion,tension=1}{photonout,leptonOneOut}       
    \end{fmfgraph*}
    \qquad\qquad
    \begin{fmfgraph*}(40,20)
      \fmfstraight
      \fmfleft{throughin,bin,dummyleftbox,dummylefttop}
      \fmflabel{$\mathrm{b}$}{bin} \fmflabel{$\bar{\mathrm{u}}$}{throughin}
      \fmfright{throughout,dout,dummyrightbox,leptonout}
      \fmflabel{$\mathrm{d}$}{dout}\fmflabel{$\ell^+$}{leptonout}\fmflabel{$\bar{\mathrm{u}}$}{throughout}
      \fmftop{dummytopveryleft,dummytopleft,leptontopout,dummytopright}
      \fmflabel{$\ell^-$}{leptontopout}
      \fmf{fermion}{throughin,throughout}
      \fmf{fermion,tension=2}{bin,bwt}
      \fmf{fermion,tension=1,label=$\mathrm{t}$}{bwt,twd}
      \fmf{fermion,tension=1}{twd,dout}
      \fmf{phantom,tension=1}{dummyleftbox,phantomWnu}
      \fmf{plain,tension=1,label=$\nu_{\ell}$,label.dist=0.2}{nuWphantom,phantomWnu}
      \fmf{phantom,tension=2}{nuWphantom,dummyrightbox}
      \fmffreeze
      \fmf{fermion,tension=1}{phantomWnu,leptontopout}
      \fmf{fermion,tension=1}{leptonout,nuWphantom}
      \fmf{photon,label=$W^-$,label.side=left,label.dist=0.01}{bwt,phantomWnu}
      \fmf{photon,label=$W^+$,label.side=left,label.dist=0.01}{twd,nuWphantom}
    \end{fmfgraph*}
  \end{center}
  \caption{\sl{The dominating $b\to d\ell\ell$ penguin graphs in the SM. Similar new physics contributions as in Fig.~\ref{fig:newPhysicsGraphs} apply.}}
  \label{fig:btodllPenguins}
\end{figure}

The smallest $B$ branching fractions measured up to now are the $b\to
s$ modes ${\cal B}(B\to K\ell\ell)=(3.4\pm0.7\pm0.2)\times10^{-7}$ and
${\cal B}(B\to
K^{*}\ell\ell)=(7.8\pm1.9\pm1.1)\times10^{-7}$.~\cite{Aubert:2006vb}
In comparison to that, ${\cal B}(B\to\pi\ell\ell)$ is expected to be
suppressed by $|V_{td}/V_{ts}|^2$, with the most recent prediction of
${\cal B}(B\to \pi\ell\ell)=3.3\times10^{-8}$.~\cite{Aliev:1998sk} Any
deviation from this order of magnitude would be a striking sign of new
physics. The two dominating diagrams for this transition are shown in
Figure~\ref{fig:btodllPenguins}, showing one penguin and one $W$ box
diagram.

\begin{figure}[t]
  \begin{center}
    \begin{minipage}{0.42\textwidth}
      \begin{center}
        \epsfig{file=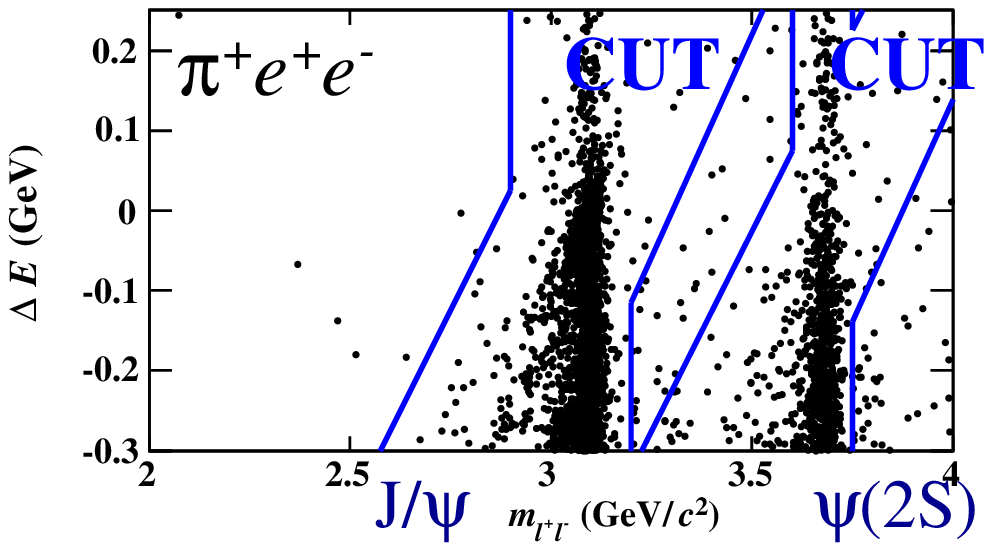, width=\textwidth}\\
        (a)
      \end{center}
    \end{minipage}
    \begin{minipage}{0.56\textwidth}
      \begin{center}
        \epsfig{file=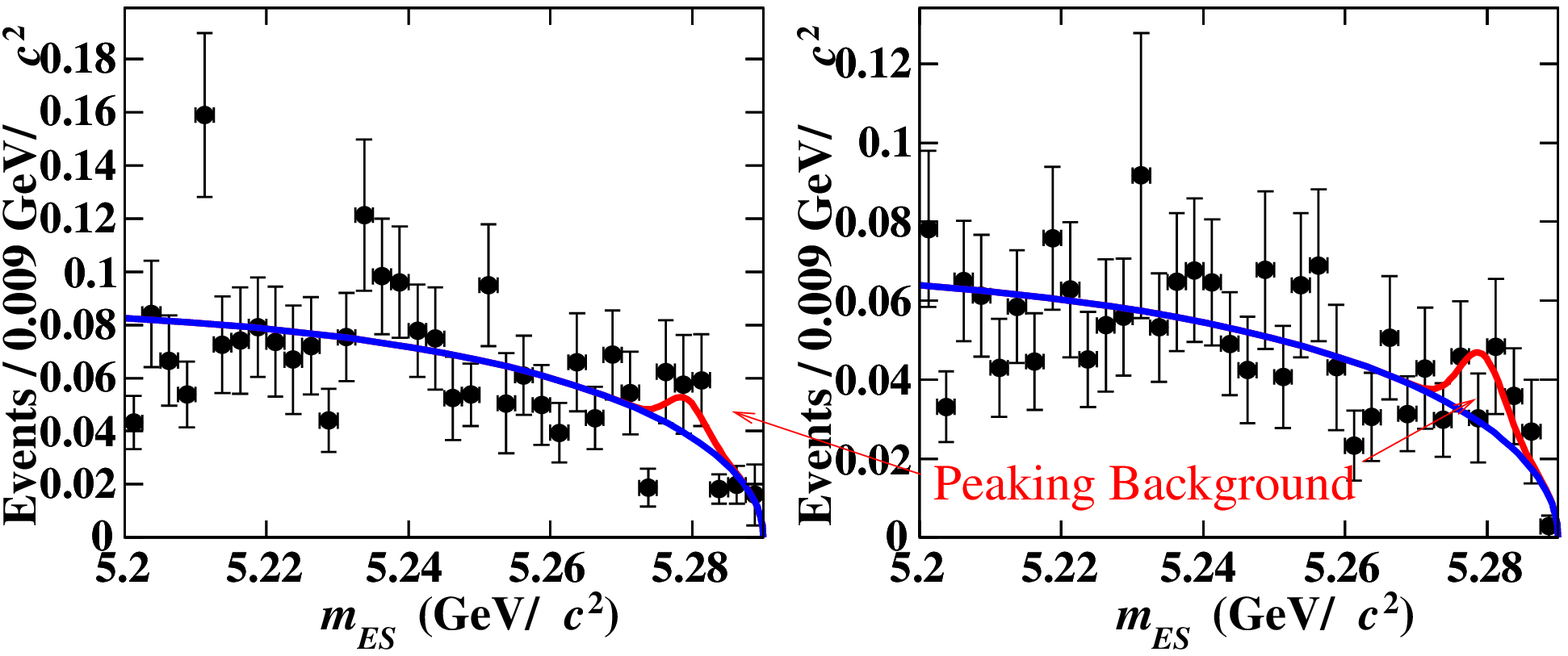, width=\textwidth}\\
        (b)
      \end{center}
    \end{minipage}
  \end{center}
  \caption{\sl{In (a), the $m_{\ell\ell},\Delta E$ distribution with
      the charmonium vetos in the $B\to \pi\ell\ell$ analysis is shown.
      (b) shows the peaking background determination in $B^0\to
      \pi^0\ell\ell$ (left) and $B^+\to \pi^+\ell\ell$ (right)}}
  \label{fig:btodllPeakingBackground}
\end{figure}

The BaBar analysis~\cite{Aubert:2007mm} on a dataset of
$209\,\mathrm{fb}^{-1}$ faces another experimental challenge with
respect to $B\to K\ell\ell$ in addition to the reduced branching
fraction, namely the much higher rate of $\pi$ in the background than
$K$. The analysis strategy is to first select clean $\pi$, $e$ and
$\mu$ candidates and then veto resonances decaying into $\ell\ell$.
The $u\bar u, d\bar d, s\bar s$ events are strongly reduced by
requiring two $p>1$~GeV leptons in the Event.  The charmonium veto
against $B\to J/\psi(\psi')\pi(K^*)$ events is shown in
Figure~\ref{fig:btodllPeakingBackground}~(a).  The tilt in the mass
veto stems from the fact that events with bremsstrahlung of one of the
leptons are off both in mass and in reconstructed energy, visible in
$\Delta E = E_{B}-1/2\,E_{Beam}$.  Event shape variables against
continuum events are grouped in a Fisher discriminant, and event shape
variables against $B\bar B$ background events are grouped into a
likelihood. After these cuts, only combinatoric background from $c\bar
c$ and $B\bar B$ events are left.

\begin{figure}[t]
  \begin{center}
         \epsfig{file=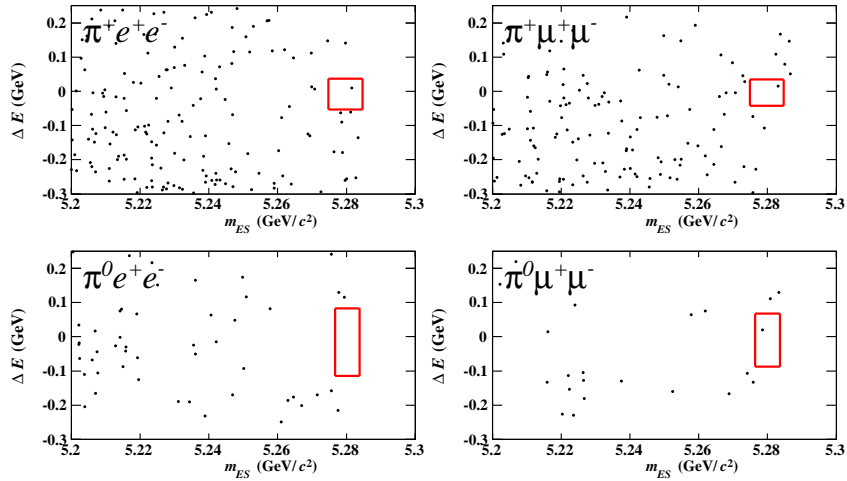, width=0.7\textwidth}
  \end{center}
  \caption{\sl{Results of the $B\to \pi\ell\ell$ analysis in the
      different modes in the $m_{ES},\Delta E$ plane. The signal boxes
      are shown in red.}}
  \label{fig:btodllResults}
\end{figure}

Backgrounds are controlled with various control samples ($e\mu$, $B\to
J/\psi(\psi')\pi(K^*)$). Hadronic mistags are measured in a separate
study by inverting the hadron vetos and then imposing the measured
mistag rates as event weights. The resulting small peaking background
fraction is shown in Figure~\ref{fig:btodllPeakingBackground}~(b) with
$\pi^+h^+h^-$ on the left and $\pi^0h^+h^-$ on the right side.
Finally, the background in the $m_{ES}$ and $\Delta E$ sidebands is
extrapolated into the signal region (see
Figure~\ref{fig:btodllResults}) to assess the background level
independent of the MC simulation.

\begin{table}[t]
  \caption{\sl{Results of the $B\to \pi\ell\ell$ selection and limits on the
      branching fraction  inferred from the absence of an excess over the SM background.}}
  \label{tab:btopillResults}
  \begin{center}
    \begin{tabular}{lccc}
      \hline\hline
      \multicolumn{2}{c}{{PRELIMINARY}}           & exp.   & BF UL \\
      Mode & obs. & backg. & 90\,\% CL $(10^{-7})$\\
      \hline
      $B^{\pm}\to \pi^{\pm}\ell\ell$ & $2$ & $1.86\pm0.38$ & $1.17$ \\
      $B^{0}\to \pi^{0}\ell\ell$     & $1$ & $0.71\pm0.30$ & $1.15$ \\
      \multicolumn{2}{l}{isospin combination}     &               & { $0.91$} \\
      $B\to \pi e\mu$                & $1$ & $2.77\pm0.70$ & $0.92$ \\
      \hline\hline
    \end{tabular}
  \end{center}
\end{table}

No excess over the background is observed, hence limits are set using
a frequentist cut-and-count method in the signal boxes. The resulting
limit of ${\cal B}(B\to \pi\ell\ell)=3.3\times10^{-8}$ assuming isospin
symmetry is within a factor of 3 of the SM prediction. A detailed
summary of the results can be found in Table~\ref{tab:btopillResults}.

\section[$b\to d\,\gamma$ Transitions]{\boldmath$b\to d\,\gamma$\unboldmath{} Transitions}\label{sec:btodg}


$b\to d\gamma$ penguins have been first observed by Belle with
$350\,\mathrm{fb}^{-1}$ with evidence for the $B^0\to\rho^0\gamma$
mode~\cite{Abe:2005rj}. In addition to the possibility of finding new
physics if the measured branching fractions exceed $\approx
0.5\times10^{-6}$ for the neutral mode or $\approx
1\times10^{-6}$ for the charged mode significantly, it offers the important possibility
to measure $|V_{td}/V_{ts}|$ via
\begin{equation}
\frac{\Gamma(B\to\rho\gamma)}{\Gamma(B\to K^*\gamma)}=\left|\frac{V_{td}}{V_{ts}}\right|^2
        \frac{(m_B-m_{\rho})^3}{(m_B-m_{K^*})^3}\left(\frac{T^{\rho}(0)}{T^{K^*}(0)}\right)^2(1+\Delta R)\vspace*{-2mm},
\label{eq:vtdvts}
\end{equation}
with $\Delta R=0.1\pm0.1$~\cite{Ali:2004hn} containing radiative
corrections and sub-dominant helicity suppressed $W$-fusion diagrams
(hence depending on $V_{ub}$ and the CKM fit itself and thus not
uncorrelated from $|V_{td}/V_{ts}|$), and a form-factor ratio of
$T^{K^*}(0)/T^{\rho}(0)=1.17\pm0.09$~\cite{Ball:2006nr}.
The graphs involved here are expected to have completely independent
possible new physics contributions than $\Delta m_{d}/\Delta m_{s}$,
hence the comparison of the two results is very important .

Here, the experimental challenge lies not only in the particle ID
requirements to suppress $K$ background, but also in the $\pi$
combinatorics coming from the wide resonance states
($\Gamma(\rho)=150\,\mathrm{MeV}$) and the high photon background from
continuum events with $\pi^0/\eta\to\gamma\gamma$. The BaBar
analysis~\cite{Aubert:2006pu} on $316\,\mathrm{fb}^{-1}$ of data
tackles this challenge by a likelihood veto against $\pi^0/\eta$ based
on the invariant mass of the photon pair and the energy of the lower
energetic photon, improving the veto significantly over a simple cut
on $m_{\gamma\gamma}$. Additionally, a Neural Net (NN) based continuum
suppression is applied. 

\begin{figure}[t]
  \begin{center}
    \epsfig{file=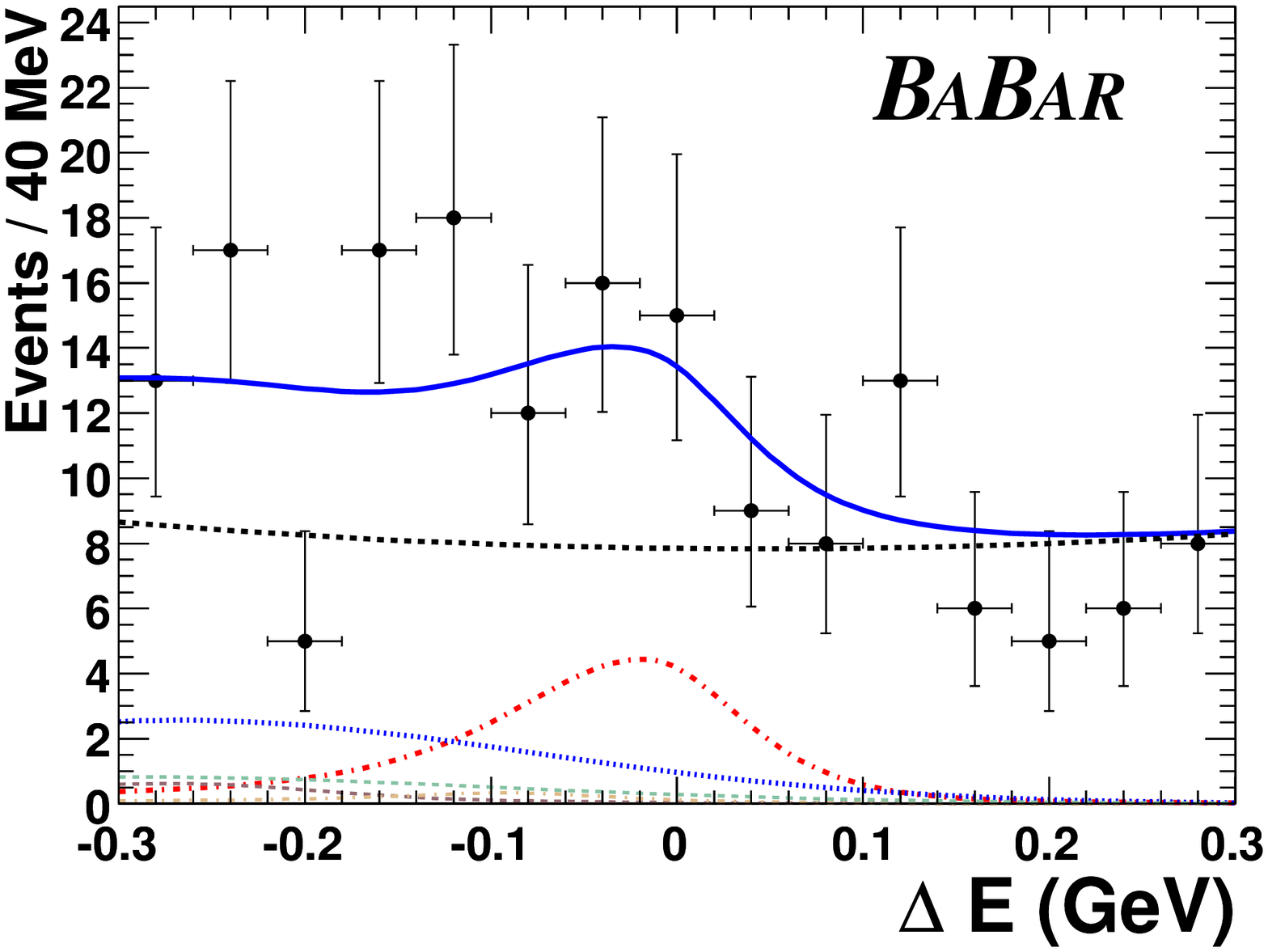, width=0.33\textwidth}
    \epsfig{file=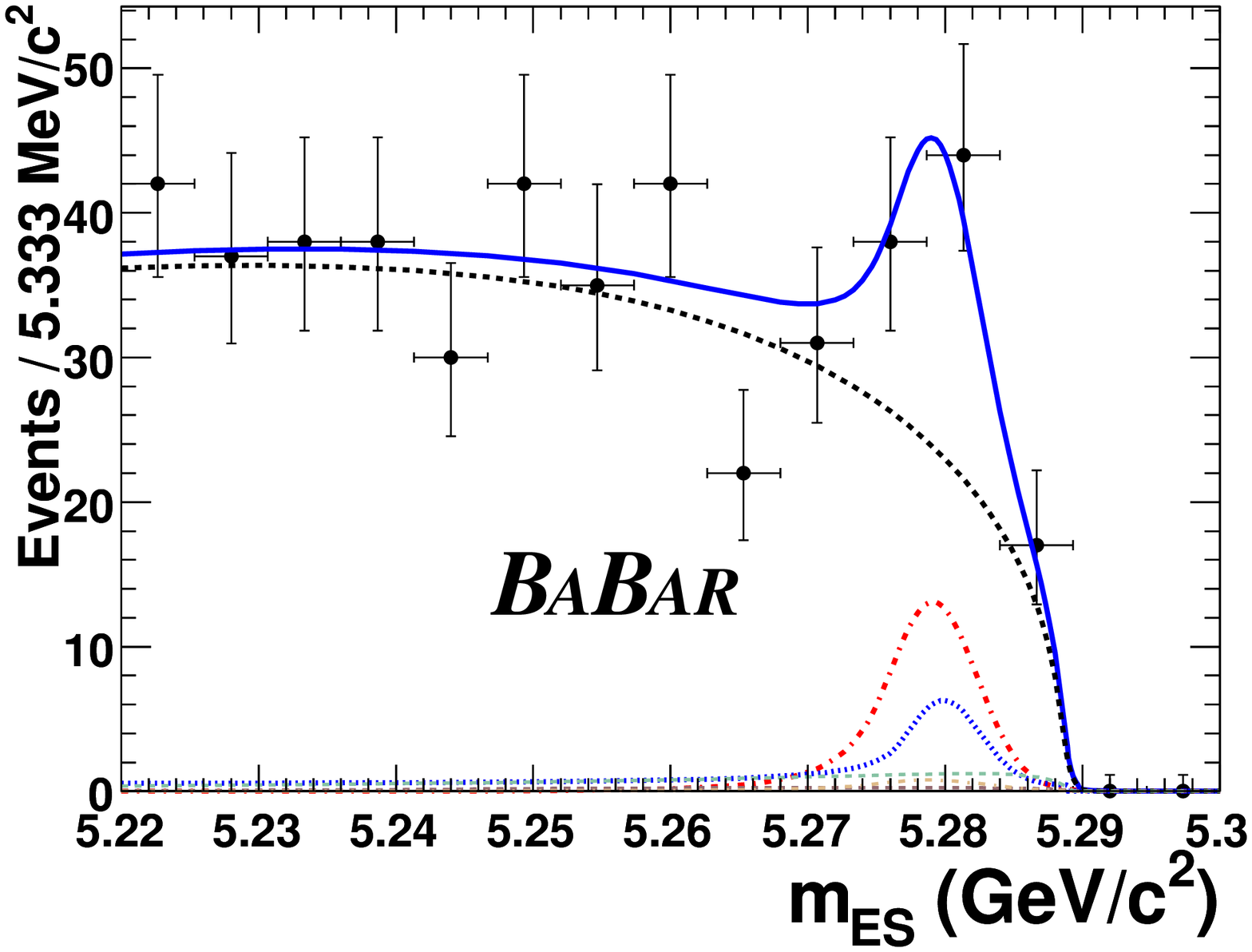, width=0.33\textwidth}
    \epsfig{file=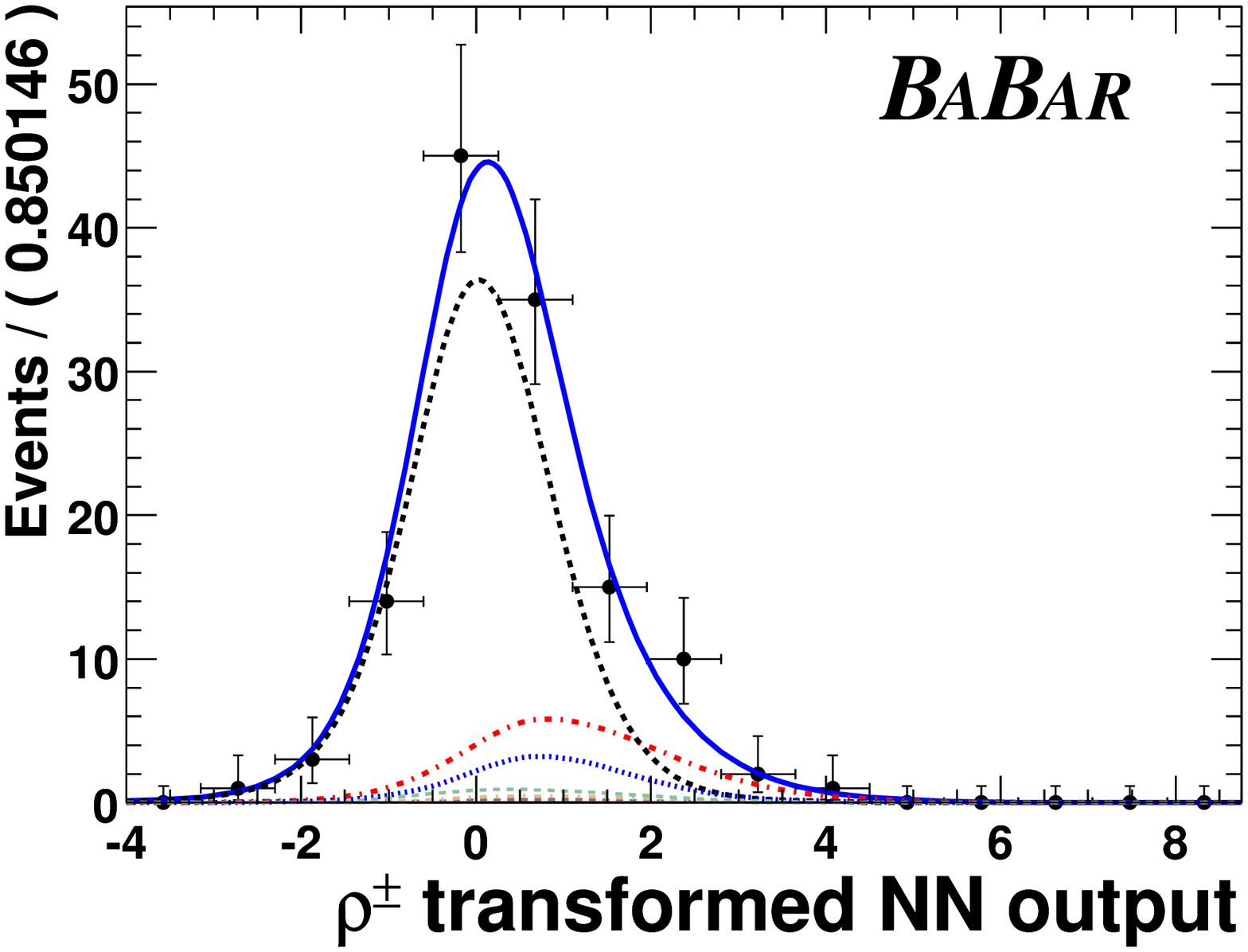, width=0.33\textwidth}
    \epsfig{file=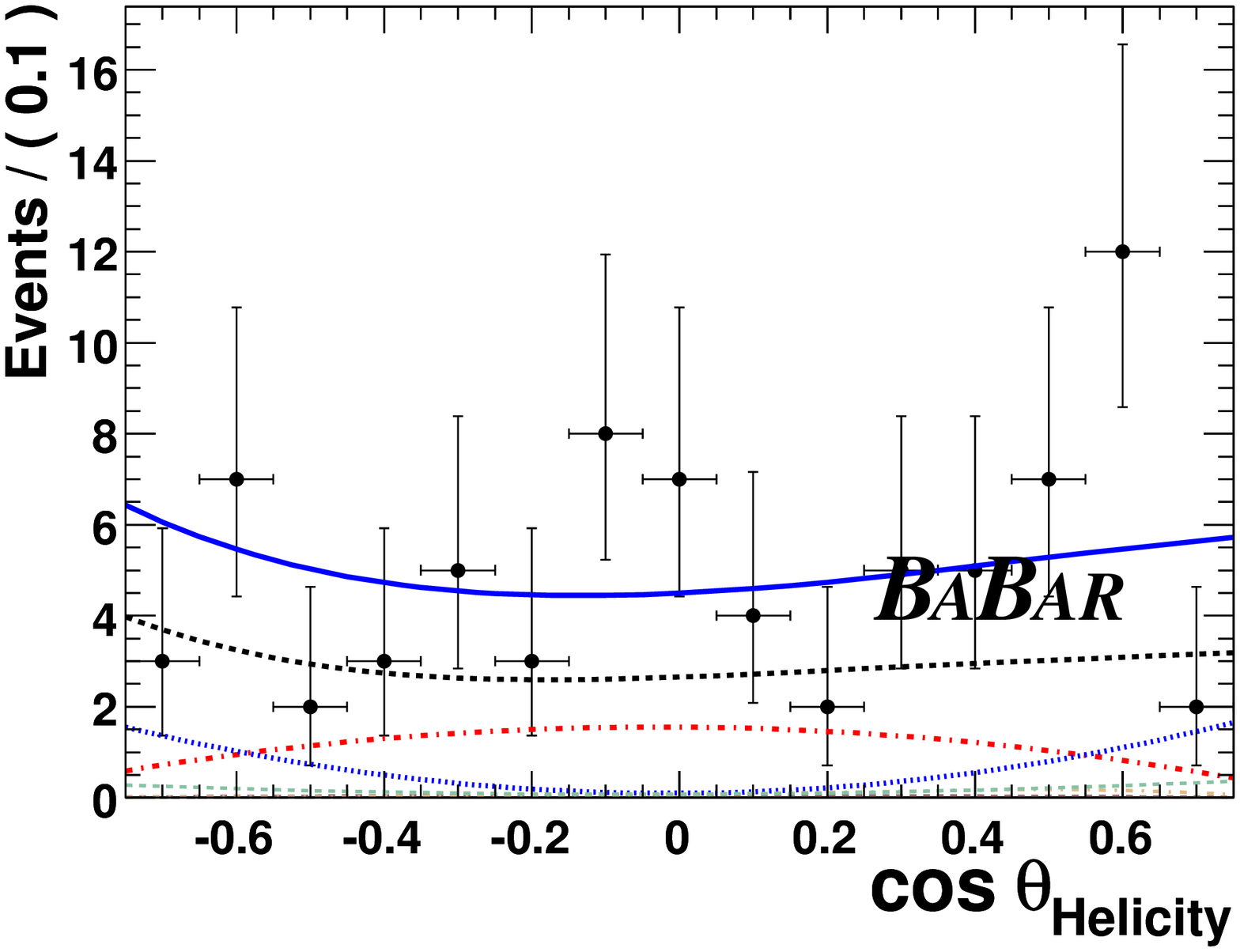, width=0.33\textwidth}\vspace{-3mm}
  \end{center}
  \caption{\sl{Extraction of the signal in the $B\to\rho/\omega\gamma$
      analysis in the $B^+\to\rho^+\gamma$ channel in a
      four-dimensional simultaneous extended maximum likelihood fit.
      The signal is shown dash-dotted (red), the continuum is
      dashed (black), the total B background is shown dotted (blue).}}
  \label{fig:rhoFit}
\end{figure}

To extract the result, a simultaneous maximum-likelihood fit is
performed to $m_{ES}, \Delta E$, the NN output and
$\cos\theta_{\mathrm{hel}}$, where $\theta_{\mathrm{hel}}$ is the
helicity angle of the vector meson $\rho$ or $\omega$, which are
transversely polarized in signal events. For $\omega$, the Dalitz
angle of the $\omega$ decay is used additionally. An example fit is
shown for the $B^+\to\rho^+\gamma$ mode in Figure~\ref{fig:rhoFit}.
Many control sample checks are performed. With off-peak data used to
control the continuum simulation, and $B\to K^*\gamma$ and $B\to D\pi$
used to control resolutions and efficiencies. 

\begin{figure}[t]
  \begin{center}
    \epsfig{file=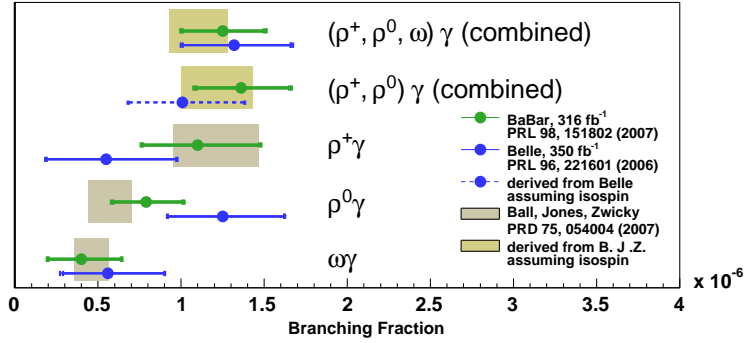, width=0.65\textwidth, clip=}\vspace{-4mm}
  \end{center}
  \caption{\sl{Graphical representation of the results of the Belle
      and Babar measurements of $b\to d\gamma$ penguins and the
      comparison with the SM prediction.}}
  \label{fig:BabarBelleComparison}
\end{figure}

\begin{table}[t]
  \caption{\sl{Results of the BaBar $B\to(\rho/\omega)\gamma$ analysis in terms of extracted branching fraction and significance. The isospin combination is shown in the last line.}}
  \label{tab:btorhogResults}
  \begin{center}
    \begin{tabular}{lcc}
      \hline\hline
      Mode & Result & Significance \\
      \hline
        ${\cal B}(B^{\pm}\to\rho^{\pm}\gamma)$ & $(1.1^{+0.37}_{-0.33}\pm0.09)\times10^{-6}$ & $3.8\,\sigma$\\
        ${\cal B}(B^{0}\to\rho^0\gamma)$ & $(0.79^{+0.22}_{-0.20}\pm0.06)\times10^{-6}$ & $4.9\,\sigma$\\
        ${\cal B}(B^{0}\to\omega\gamma)$ & $(0.40^{+0.24}_{-0.20}\pm0.05)\times10^{-6}$ & $2.2\,\sigma$\\
        \hline
\multicolumn{3}{c}{combination of all modes}\\
        ${\cal B}(B\to\rho/\omega\gamma)$ & $(1.25^{+0.25}_{-0.24}\pm0.09)\times10^{-6}$ & $6.4\,\sigma$ \\
      \hline\hline
    \end{tabular}
  \end{center}
\end{table}

The result for the individual modes and different combinations is
shown in Table~\ref{tab:btorhogResults}. While the
$B^{0}\to\omega\gamma$ signal is not yet significant on its own, this
result represents the first evidence for $B^{+}\to\rho^+\gamma$.
Figure~\ref{fig:BabarBelleComparison} shows the good agreement of the
results with the SM predictions and the agreement with the earlier
Belle results. The result for the isospin combination
$B\to\rho/\omega\gamma$ is to be taken with care, however, since the
$\omega$ does not belong to the isospin triplet.


\begin{figure}[t]
  \begin{center}
    \epsfig{file=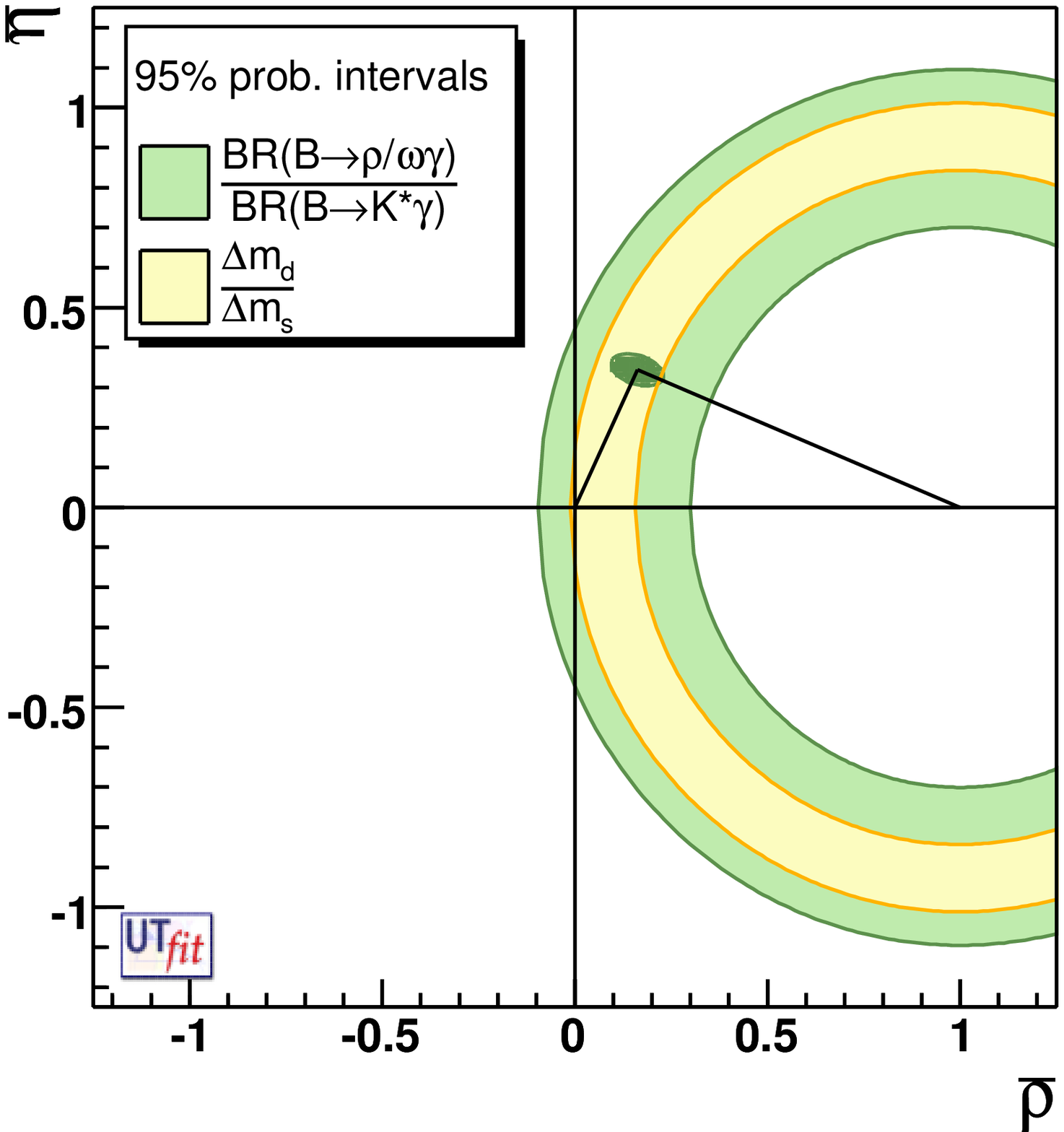, width=0.5\textwidth,clip=}
  \end{center}
  \caption{\sl{The constraint on $|V_{td}/V_{ts}|$ from
      $B\to\rho/\omega\gamma$ and $\Delta m_d/\Delta m_s$ in the
      $\bar\rho,\bar\eta$ plane and the comparison with the SM CKM
      fit.}}
  \label{fig:DeltaMvsBtoRho}
\end{figure}

The results of ${\cal B}(B\to\rho/\omega\gamma)$ can be
used to extract $|V_{td}/V_{ts}|$ using Eq.~\ref{eq:vtdvts}. Using the
world average of ${\cal
  B}(B\to\rho/\omega\gamma)=(1.25^{+0.25}_{-0.24}\pm0.09)\times10^{-6}$,
a value of
$|{V_{td}}/{V_{ts}}|_{\rho/\omega\gamma}=0.202^{+0.017}_{-0.016}
\pm0.015$~\cite{Ball:2006eu} can be extracted, in very good agreement
with the result obtained from $\Delta m_d/\Delta m_s$ of
$|{V_{td}}/{V_{ts}}|_{\Delta m_d/\Delta
  m_s}=0.2060\pm0.0007^{+0.0081}_{-0.0060}$ from the
Tevatron~\cite{Abulencia:2006ze}. While the precision from
$B\to\rho/\omega\gamma$ is not sufficient to compete with $\Delta
m_d/\Delta m_s$, it provides an important independent check of
possible new physics, due to the different nature of the possible new
physics contributions in the two processes. A comparison of the two
results with the CKM fit and the imposed constrained in the
$\bar\rho,\bar\eta$ plane of the CKM parameterization can be found in
Fig.~\ref{fig:DeltaMvsBtoRho}.

\section{Summary and Outlook}\label{sec:summary}

\begin{figure}[t]
  \begin{center}
    \epsfig{file=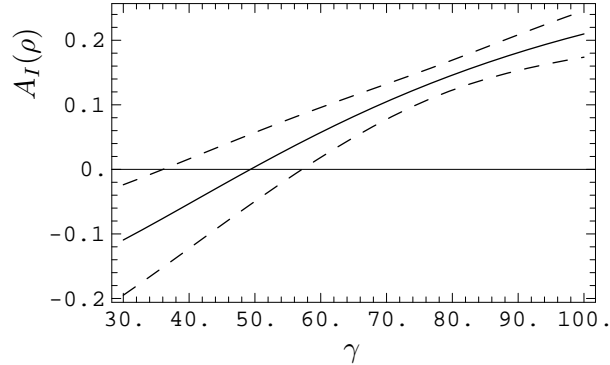, width=0.5\textwidth}
  \end{center}
  \caption{\sl{A possible new constraint on the CKM-angle $\gamma$
      from the isospin asymmetry measurement in $B\to\rho\gamma$.
    }}
  \label{fig:gammaAI}
\end{figure}

With the extraordinary luminosities of the B-factories Belle and
BaBar, the field of strange radiative penguin decays of the B meson
has evolved into one of the most important areas of precision
measurements in the search of new physics. While it is still unclear
whether $B\to\pi\ell\ell$ transitions will be seen in the lifetime of
the present B-factories, the rare decay $b\to d\gamma$ has now been
seen by Belle and BaBar, and these measurements will possibly evolve
towards precision measurements in the same way as the $b\to s\gamma$
decays before. With the anticipated ${\cal L}_{\mathrm{int}}\approx
1\,\mathrm{ab}^{-1}$ of luminosity per B-factory at the end of 2008, a
$\approx10\,\%$ measurement of the CP asymmetry in
$B\to\rho/\omega\gamma$ should be feasible. Another interesting
measurement would be the isospin asymmetry
$A_I={2\Gamma(B^0\to\rho^0\gamma)}/{\Gamma(B^{\pm}\to\rho^{\pm}\gamma)}-1$,
which presents a completely new way of obtaining a measurement of the
CKM-angle $\gamma$, as outlined in~\cite{Ball:2006eu} and shown in
Fig.~\ref{fig:gammaAI}.


\section*{Acknowledgments}

The author would like to thank the PEP-II accelerator and its crew for
the excellent luminosity delivered to BaBar, and the radiative penguin
analysis group in BaBar for lots of support and very lively discussions.

\end{fmffile}

\section*{References}
\begin{thebibliography}{99}

\bibitem{Ammar:1993sh}
  R.~Ammar {\it et al.}  [CLEO Collaboration],
  Phys.\ Rev.\ Lett.\  {\bf 71} (1993) 674.

\bibitem{Misiak:2006zs}
  see e.g.~M.~Misiak {\it et al.},
  Phys.\ Rev.\ Lett.\  {\bf 98} (2007) 022002
  and references therein.

\bibitem{Aubert:2006vb}
  B.~Aubert {\it et al.}  [BABAR Collaboration],
  Phys.\ Rev.\  D {\bf 73} (2006) 092001.

\bibitem{Aliev:1998sk}
  T.~M.~Aliev and M.~Savci,
  Phys.\ Rev.\  D {\bf 60}, 014005 (1999).

\bibitem{Aubert:2007mm}
  B.~Aubert {\it et al.}  [BABAR Collaboration],
  arXiv:hep-ex/0703018, submitted to PRL.

\bibitem{Abe:2005rj}
  K.~Abe {\it et al.}  [BELLE Collaboration],
  Phys.\ Rev.\ Lett.\  {\bf 96} (2006) 221601.

\bibitem{Ali:2004hn}
  A.~Ali, E.~Lunghi and A.~Y.~Parkhomenko,
  Phys.\ Lett.\  B {\bf 595}, 323 (2004).

\bibitem{Ball:2006nr}
  P.~Ball and R.~Zwicky,
  JHEP {\bf 0604}, 046 (2006).

\bibitem{Aubert:2006pu}
  B.~Aubert {\it et al.}  [BABAR Collaboration],
  Phys.\ Rev.\ Lett.\  {\bf 98} (2007) 151802.

\bibitem{Ball:2006eu}
  P.~Ball, G.~W.~Jones and R.~Zwicky,
  Phys.\ Rev.\  D {\bf 75} (2007) 054004.

\bibitem{Abulencia:2006ze}
  A.~Abulencia {\it et al.}  [CDF Collaboration],
  Phys.\ Rev.\ Lett.\  {\bf 97}, 242003 (2006).

\end{thebibliography}
\end{document}